\documentclass[a4paper]{jpconf}
\usepackage{graphicx,epsfig,wrapfig}
\begin{document}
\title{Plans for Hadronic Structure Studies at J-PARC}
\author{S. Kumano}
\address{KEK Theory Center, Institute of Particle and Nuclear Studies, \\
             High Energy Accelerator Research Organization (KEK) \\
             and Department of Particle and Nuclear Studies,
             Graduate University for Advanced Studies \\
             1-1, Ooho, Tsukuba, Ibaraki, 305-0801, Japan}
\ead{shunzo.kumano@kek.jp}
\begin{abstract}
Hadron-physics projects at J-PARC are explained. 
The J-PARC is the most-intense hadron-beam facility
in the multi-GeV high-energy region. By using secondary beams 
of kaons, pions, and others as well as the primary-beam proton,
various hadron projects are planned.
First, some of approved experiments are introduced
on strangeness hadron physics and hadron-mass modifications
in nuclear medium. Second, future possibilities are discussed 
on hadron-structure physics, including structure functions
of hadrons, spin physics, and high-energy hadron reactions
in nuclear medium. The second part is discussed in more details
because this is an article in the hadron-structure session.
\end{abstract}

\vspace{-0.6cm}
\section{Introduction}\label{intro}

\begin{figure}[t]
\begin{minipage}{0.48\textwidth}
\includegraphics[width=1.00\textwidth]{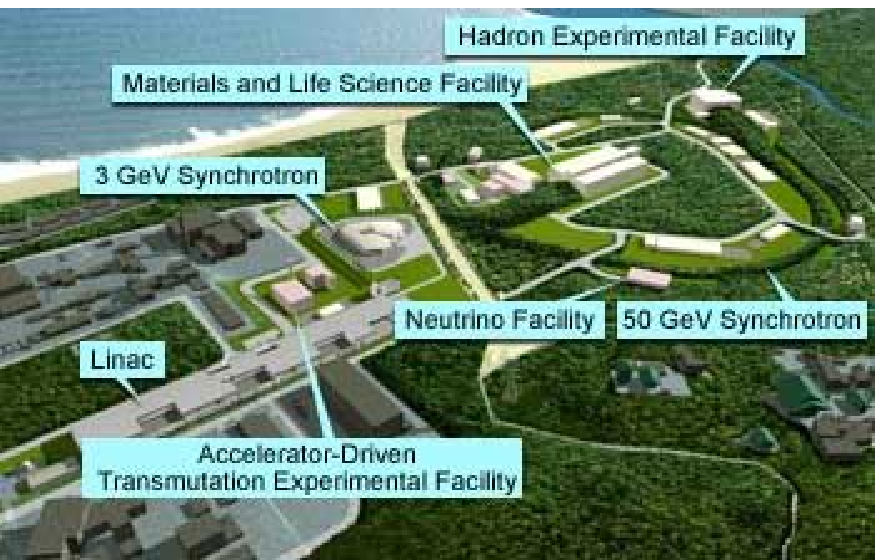}
\caption{J-PARC facility \cite{j-parc}.}
\label{fig:j-parc}
\end{minipage}
\hspace{0.5cm}
\begin{minipage}{0.48\textwidth}
\includegraphics[width=1.00\textwidth]{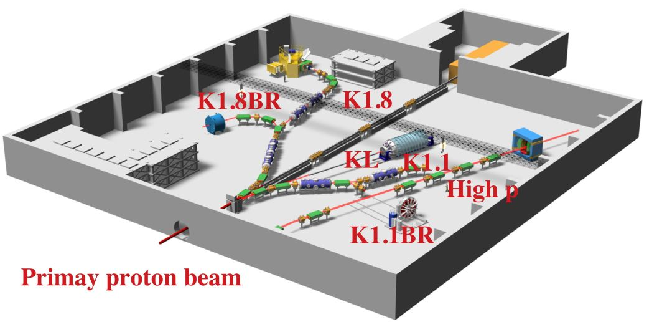}
\caption{Beamline layout of hadron hall.}
\label{fig:hadron-hall}
\end{minipage} 
\end{figure}

The J-PARC stands for Japan Proton Accelerator Research Complex
\cite{j-parc}, which is located at Tokai in Japan. It is a multi-purpose
facility ranging from life sciences to nuclear and particle physics
\cite{inpc10-nagamiya-nagae,sawada}.
The J-PARC accelerator consists of a linac as an injector,
a 3-GeV rapid cycling synchrotron, and a 50 GeV synchrotron
as shown in Fig. \ref{fig:j-parc}.
The J-PARC has three major projects:
(1) material and life sciences with neutrons and
muons produced by the 3-GeV proton beam,
(2) nuclear and particle physics with secondary beams 
as well as the primary-proton beam, 
(3) nuclear transmutation by the linac.
The advantage of the facility is the beam intensity:
1 MW in the 3-GeV synchrotron and 0.75 MW in the 50 GeV one.
At this beginning stage, the proton is accelerated to 30 GeV instead of
the original 50 GeV, and the beam intensity has not
reached to the designed level.

Hadron experiments are planned in the hadron experimental facility
in Figs. \ref{fig:j-parc} and \ref{fig:hadron-hall}. 
Various secondary beams such as pions, kaons, and anti-proton
as well as the primary-proton beam are used for 
hadron-physics experiments \cite{j-parc,inpc10-nagamiya-nagae,sawada}. 
The K1.8 and K1.1 beamlines are mainly for kaon beams
with the momentum around 1.8 GeV/$c$ and 1.1 GeV/$c$, respectively.
The KL beamline is for measurements of neutral kaon decays, and
the high-momentum beamline is for the primary proton beam.
Currently, the K1.8, K1.8BR, KL, and K1.1BR beamlines are ready,
the K1.1 will be ready in the near future, and the high-momentum
beamline needs to be constructed.

In this article, the ``hadron physics" is used by including 
nuclear projects in addition to hadron projects in a narrow sense.
The hadron physics could also include neutrino interactions 
with nuclei in the T2K neutrino-oscillation experiment. 
However, this topic is not discussed in this article, so that
one may look at discussions in a dedicated international workshop
on the neutrino-nucleus interactions \cite{nuint09}.
The hadron-physics experiments at the hadron hall start from
strangeness hadron physics such as hypernuclear physics
and a test of pentaquark existence. Then, nuclear medium effects
on hadron masses will be investigated. There are proposals
on dimuon production ($J/\psi$, Drell-Yan); however, they
are not approved at this stage. 
In this article, we first explain some of the approved projects in 
Sec. \ref{approved}, and then possible hadron-structure projects
are introduced in Sec. \ref{structure}. 
Here, we focus on hadron-structure projects
because this article is in the hadron-structure session.
These hadron-physics studies at J-PARC are summarized 
in Sec. \ref{summary}.

\section{Approved projects}\label{approved}

All the J-PARC proposals on hadron physics are listed in 
Ref. \cite{j-parc-proposal}. Among them, some of
approved experiments are explained first in this section.
We list major purposes of these projects:
\begin{itemize}
\vspace{-0.14cm}
\item[(1)] creation of new forms of hadronic systems by extending
           flavor degrees of freedom,
\vspace{-0.07cm}
\item[(2)] studies of hyperon interactions and their applications
           to neutron stars,
\vspace{-0.07cm}
\item[(3)] search of exotic hadrons including strangeness,
\vspace{-0.07cm}
\item[(4)] investigations on hadron-mass generation mechanism.
\end{itemize}
\vspace{-0.10cm}
These projects should open a new realm of hadron physics.

\subsection{Hypernuclear physics}\label{hyper}

\begin{wrapfigure}{r}{0.48\textwidth}
   \vspace{-1.1cm}
\begin{center}
   \includegraphics[width=0.45\textwidth]{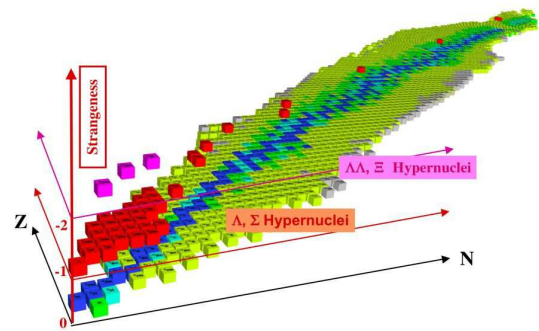}
\end{center}
\vspace{-0.4cm}
\caption{Nuclear chart with strangeness \cite{kaneta}.}
\label{fig:n-chart}
\end{wrapfigure}

In the beginning stage, one of the major projects is the hypernuclear
physics \cite{inpc10-nagamiya-nagae}.
It is intended to find new forms of hadronic many-body systems
by extending flavor degrees of freedom as shown in 
Fig. \ref{fig:n-chart}, where the strangeness is taken as the third axis. 
The strangeness hadron physics is interesting with the following reasons.
First, the strange-quark mass is the order of the QCD scale parameter 
$\Lambda_{QCD}$, which suggests that the strangeness is an appropriate
quantity to probe QCD dynamics.
Second, there is no Pauli blocking for hyperons ($Y$) in a nucleus,
so that they are good probes of deep regions of nuclei. 
Third, hyperons could exist in neutron stars, so that information
on $YN$ and $YY$ interactions should be important for understanding
their properties \cite{neutron-star}.

As shown in Fig. \ref{fig:n-chart}, there are some data for hypernuclei
with $S=-1$ but there are only a few data in the double strangeness plane.
It is one of the major purposes of the J-PARC projects to find many new
hypernuclei especially with $S=-2$. Then, $YY$ and $YN$ interactions
should be clarified from their properties. Independent and
direct measurements of $YN$ interactions are also considered, but
there is no actual proposal at this stage. 
These interactions have renewed interests because of recent developments
in the field of lattice QCD for nuclear force and light nuclei
\cite{lattice-nn}. 

\subsection{Exotic hadrons}\label{strange}

Exotic hadrons indicate hadrons with internal configuration
other than ordinary $q\bar q$ and $qqq$ types.
This topic has been investigated for a long time since the early stage
of quark models; however, an undoubted experimental evidence has not been
found yet. Nonetheless, there are recent discoveries of their candidates
in charmed hadrons particularly from the Belle and BaBar collaborations. 
In this section, kaonic nuclei are included as one of such projects.
It seems that time has come to establish the field of exotic hadrons
with experimental confirmations. In Japan, an exotic-hadron project
started in 2009 by forming groups of many experimentalists and
theorists \cite{exotic-project} for searching exotics at KEKB, SPring8,
and J-PARC, and its activities are in progress.

\begin{wrapfigure}{r}{0.38\textwidth}
   \vspace{-0.6cm}
\begin{center}
   \includegraphics[width=0.36\textwidth]{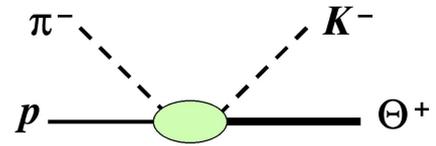}
\end{center}
    \vspace{-0.2cm}
\caption{Test of $\Theta^+$ existence.}
\label{fig:theta}
\end{wrapfigure}

The first experiment at the K1.8 beamline is a test of $\Theta^+$
existence by the $\pi^- p \rightarrow K^- \Theta^+$ reaction
as shown in Fig. \ref{fig:theta}.
A possible pentaquark state $\Theta^+$ has been controversial
for several years due to many negative experiments.
On the other hand, there are still positive measurements.
For example, a recent report from the LEPS collaboration
repeatedly indicated its existence \cite{theta}.
It is important to do a decisive measurement at J-PARC.
An $S$-channel formation by $K^+ + d$ should be such an experiment
\cite{s-theta}.
However, the reaction $\pi^- p \rightarrow K^- \Theta^+$ 
will be investigated first because the kaon-beam intensity is 
still low in the beginning operation of the J-PARC.

\begin{figure}[b]
\begin{minipage}{0.48\textwidth}
   \vspace{-0.4cm}
\begin{center}
   \includegraphics[width=0.90\textwidth]{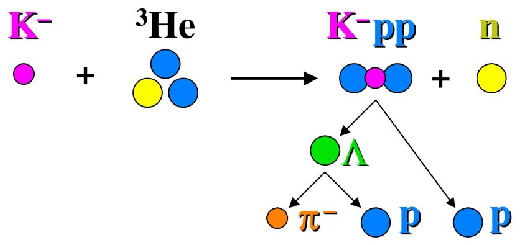}
\end{center}
\vspace{-0.4cm}
\caption{Search for $K^-pp$ bound state \cite{kpp}.}
\label{fig:kpp}
\end{minipage}
\hspace{0.5cm}
\begin{minipage}{0.48\textwidth}
   \vspace{-0.4cm}
\begin{center}
   \includegraphics[width=0.85\textwidth]{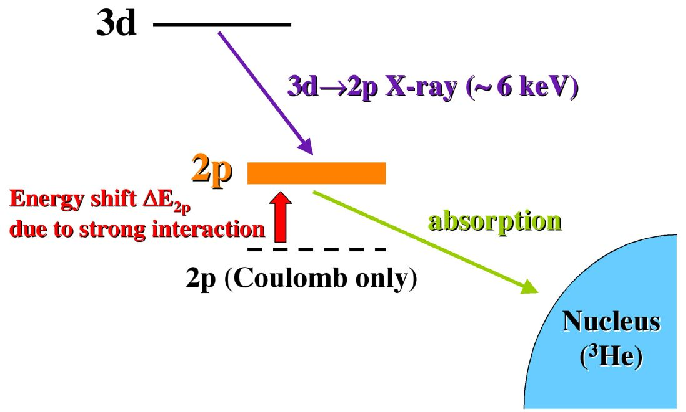}
\end{center}
   \vspace{-0.4cm}
\caption{X ray from kaonic helium-3 \cite{k-he-x}.}
\label{fig:khe-x}
\end{minipage} 
\end{figure}

Few-body bound states of kaon and nucleons are also interesting
since there are experimental reports on their candidates.
This topic originates from an indication that $\Lambda(1405)$
is a bound state of $\bar K N$. Then, an attractive $\bar K N$
interaction could make it possible to form bound states
$\bar K NN$, $\bar K NNN$, and so on \cite{knnn}, which are 
called kaonic nuclei. Recent research activities are focused on
simple systems $\Lambda(1405)$ and $K^-pp$ among them.
The first experiments at the K1.8BR beamline are on 
the $K^- pp$ bound state and $\bar K N$ interactions 
as shown in Figs. \ref{fig:kpp} and \ref{fig:khe-x}, respectively.
The $K^-pp$ bound state is searched in two ways by
missing-mass spectroscopy with the measurement of a neutron
and by invariant-mass reconstruction with the measurement
of decay particles. In the E17 experiment in Fig. \ref{fig:khe-x},
the $X$-ray is observed for the $3d \rightarrow 2p$ transition of
the kaonic helium-3. The $2p$ energy level is sensitive to
the strong interaction of $\bar K N$, and its effect should be
reflected in the $X$-ray energy. Therefore, this experiment
provides us valuable information for finding the controversial
strength of $\bar K N$ interactions. There are also experiments
for investigating the nature of $\Lambda (1405)$ \cite{lambda-1405}.

\subsection{Hadron masses in nuclear medium}\label{medium}

\begin{wrapfigure}{r}{0.36\textwidth}
   \vspace{-0.8cm}
\begin{center}
   \includegraphics[width=0.28\textwidth]{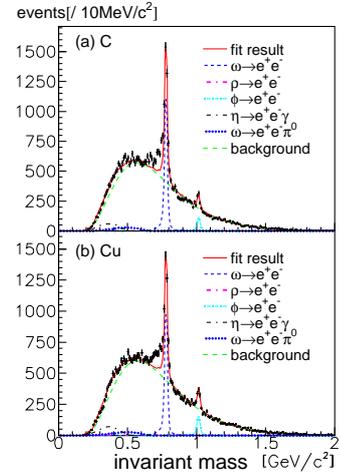}
\end{center}
    \vspace{-0.6cm}
\caption{Meson-mass measurements in nuclear medium \cite{kek-ps}.}
\label{fig:meson-masses}
\end{wrapfigure}

The origin of hadron masses will be investigated by using
the primary-proton beamline by observing light vector-meson masses
in nuclear medium. We know that up- and down-quark masses are much
smaller than the nucleon mass, which makes us wonder
how the major part of the nucleon mass, or generally hadron masses,
is generated. A possible idea is due to chiral symmetry breaking.
An order parameter of the breading is the quark condensate $<q\bar q>$.
However, it is not an observable, so that meson-mass shifts,
which are connected to the condensate, are investigated
experimentally in nuclear medium \cite{hh-summary}.

This topic has been investigated in various facilities including
the KES-PS with the primary 12-GeV proton beam:
$p+A \rightarrow V +X$ 
($V=\rho, \ \omega, \ \phi \ \rightarrow e^+e^-$) \cite{kek-ps}
as shown in Fig. \ref{fig:meson-masses}.
They observed 9\% and 3\% mass shifts for $\omega$ ($\rho$) 
and $\phi$, respectively. These measurements will be continued 
at J-PARC with much better statistics.

\vspace{-0.1cm}
\section{Possibilities of hadron structure studies}\label{structure}

Since this article is in the hadron-structure session, it is 
appropriate to focus on hadron-structure projects at J-PARC.
Some ideas are introduced in this section; however, the reader should 
aware that the following topics have not been officially approved yet 
or even a proposal does not exist for some projects.
It inevitably means that discussed topics are somewhat based on
author's personal view. 
New ideas are welcome as the forms of experimental
proposals and/or contributions to J-PARC workshops \cite{j-parc-workshop}.
All the following projects, as well as the above meson-mass measurements,
use the high-momentum primary-proton beamline as indicated ``High p" 
in Fig. \ref{fig:hadron-hall}. This beamline needs to be constructed
in the near future.
We list major purposes of the hadron-structure projects:
\begin{itemize}
\vspace{-0.14cm}
\item[(1)] investigations on applicability of perturbative QCD,
\vspace{-0.07cm}
\item[(2)] search for origin of nucleon spin,
\vspace{-0.07cm}
\item[(3)] establishment of parton distribution functions at large $x$,
\vspace{-0.07cm}
\item[(4)] investigations on mechanisms of quark and hadron 
           interactions in nuclear medium.
\end{itemize}
\vspace{-0.1cm}
These studies lead to the establishment of hadron and nuclear 
structure at large $x$ as a complimentary project to RHIC and LHC.
They also provide basic information on proton structure 
for new discoveries such as at LHC.

\vspace{-0.2cm}
\subsection{Applicability of perturbative QCD}
\label{pqcd}

The proton-beam energy of 50 GeV (currently 30 GeV)
corresponds to the c.m. energy $\sqrt{s}=10$ GeV, which is 
rather low in comparison with the RHIC and LHC energies.
It suggests that perturbative QCD (pQCD) 
corrections should be large.
In order to extract meaningful results on parton structure,
one needs to remove perturbative effects from measured data.
Therefore, it is essential to understand the pQCD corrections
at the J-PARC energy. 

In the last several years, theoretical techniques, especially
on gluon resummations, have been developed for describing
fixed-target cross sections. It is now known that the large corrections
mainly come from threshold gluon emissions 
($m_{\mu\mu}^2 \sim \hat s_{q\bar q}$ in Drell-Yan processes
with dimuon mass $m_{\mu\mu}$ and c.m. energy squared $s_{q\bar q}$
for $q \bar q$). Such corrections are estimated for the Drell-Yan
processes at the J-PARC energy \cite{resum}, and the results 
indicate that the resummation effects are indeed large. 
However, NLL (next-to-leading logarithmic) and NNLL 
(next-to-next-to-leading logarithmic) cross sections are
similar, which indicates that the resummation series are
converging in the Drell-Yan if the NNLL resummations are
taken at $\sqrt{s}=10$ GeV.

The J-PARC energy is the transition region from hadron degrees
of freedom (d.o.f.) to quark-gluon d.o.f., and it is a boundary 
in applying perturbative QCD. For the Drell-Yan at $\sqrt{s}=10$ GeV,
the pQCD corrections can be understood by including the gluon resummations,
so that parton-structure information should be extracted from the data.
On the other hand, this energy region is challenging and attractive
for pQCD theorists in testing their calculations.
Such basic studies are essential for establishing hadron physics 
from a description in terms of hadrons to the one in terms of
quarks and gluons.

\subsection{Flavor dependence of antiquark distributions}
\label{antiquark}

\begin{wrapfigure}{r}{0.36\textwidth}
   \vspace{-1.0cm}
\begin{center}
   \includegraphics[width=0.28\textwidth]{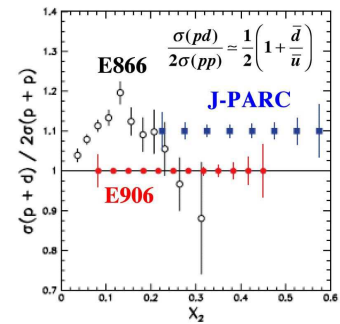}
\end{center}
    \vspace{-0.6cm}
\caption{Drell-Yan cross-section ratios \cite{p04}.}
\label{fig:ub-db}
\end{wrapfigure}

There are proposals on dimuon experiments by using the primary proton
beam \cite{p04,p24}. The J/$\psi$ production and Drell-Yan measurements
are intended to investigate structure functions at medium Bjorken
variable $x$.
The E906/SeaQuest experiment at Fermilab is in progress for dimuon
experiments, and it could be continued at J-PARC.
For example, the Drell-Yan cross sections are measured for 
$pp$ and $pd$ ($p$: proton, $d$: deuteron), and then their ratio 
is taken for finding the flavor dependence of antiquark distributions:
$\sigma_{DY}^{pd}/(2 \sigma_{DY}^{pp}) \approx (1+\bar d/\bar u)/2$
as shown in Fig. \ref{fig:ub-db}.
The Fermilab-E866 experiment provided a clear evidence for 
the flavor-asymmetric antiquark distributions $\bar u \ne \bar d$.
The E906/SeaQuest and J-PARC measurements are expected to extend the $x$
range to a larger-$x$ region. In particular, the measured E866 ratios
tend to decrease as $x$ becomes larger in Fig. \ref{fig:ub-db},
which is difficult to be interpreted theoretically.
The flavor asymmetric antiquark distributions cannot be explained
by a perturbative QCD mechanism because up- and down-quark masses
are very small, so that it is mainly associated with nonperturbative
properties such as pion clouds in the nucleon \cite{flavor}.

Measurements of antiquark distributions in nuclei are also
interesting. The Fermilab Drell-Yan measurements showed
that nuclear modifications of the antiquark distributions are
very small at $x \sim 0.1$, which ruled out the pion-excess mechanism.
Since the past Fermilab measurements are limited to a narrow
kinematical region at $x \sim 0.1$, it is desirable to extend
them to larger $x$ by J-PARC experiments.
Determination of nuclear parton distribution functions (PDFs)
at large $x$ is valuable for understanding the nuclear
modification mechanism and in general for precisely calculating
other high-energy reactions, for example, high-$p_T$ jet and 
hadron production cross sections at LHC.

\subsection{Spin physics without proton-beam polarization}\label{spin-unpol}

$ \ \ \ $ \vspace{-0.4cm}

\begin{wrapfigure}{r}{0.50\textwidth}
   \vspace{-0.4cm}
\begin{center}
   \includegraphics[width=0.48\textwidth]{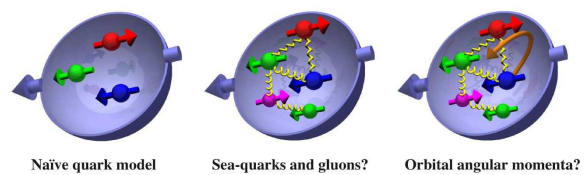}
\end{center}
    \vspace{-0.3cm}
\caption{Origin of nucleon spin \cite{hermes}.}
\label{fig:n-spin}
\end{wrapfigure}

\noindent
The nucleon spin is one of fundamental physics quantities, 
yet its origin is not known in terms of quark and gluon degrees
of freedom. In a simple quark model, it is supposed to be explained
by a three-quark-spin combination as shown in the left part 
of Fig. \ref{fig:n-spin},
which is denied by polarized lepton-nucleon scattering
experiments \cite{pol-pdfs}.
The gluon polarization seems to be also small according to
recent lepton-scattering and RHIC-Spin measurements, so that
the remaining possibility is a significant contribution from
orbital angular momenta as illustrated in Fig. \ref{fig:n-spin}.
The J-PARC facility could contribute various aspects of such 
high-energy spin studies.
Since the current proton-beam energy is 30 GeV without polarization,
initial spin-physics projects should be carefully
planned with target polarizations and new observables, which are
not (well) investigated at the Brookhaven AGS.

\vspace{-0.3cm}
\subsubsection{
Transverse-momentum-dependent distributions \\}

$ \ \ \ $ \vspace{-0.4cm}

\begin{wrapfigure}{r}{0.55\textwidth}
   \vspace{-0.4cm}
\begin{center}
   \includegraphics[width=0.52\textwidth]{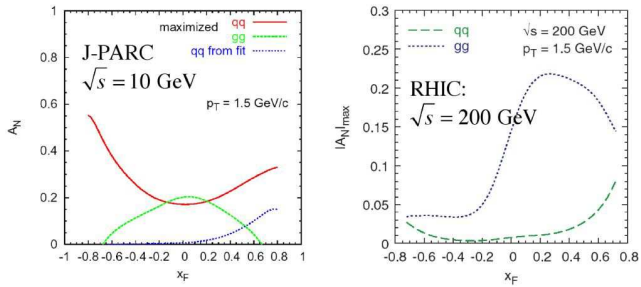}
\end{center}
    \vspace{-0.6cm}
\caption{Single spin asymmetry in $D$-meson production
at J-PARC and RHIC \cite{p24,j-parc-d}.}
\label{fig:ssa-d}
\end{wrapfigure}

\noindent
Nucleon structure has been investigated in a collinear factorization
form by integrating transverse momenta; however, it is now becoming
possible to investigate more details on transverse structure.
Such studies are valuable for clarifying the nucleon spin issue,
particularly on the orbital-angular-momentum part,
and in general for describing high-energy hadron reactions at RHIC
and LHC.

Aforementioned Drell-Yan measurements can be used for investigating
a transverse-momentum-dependent (TMD) polarized parton distributions,
so called Boer-Mulders (BM) functions, by observing violation of
the Lam-Tung relation in the Drell-Yan cross sections \cite{p24}.
The BM functions indicate transversely-polarized quark distributions
in the unpolarized nucleon. 
Other interesting TMD distributions are the Sivers functions,
which indicate unpolarized quark distributions in 
the transversely-polarized nucleon. For example, they appear
in single spin asymmetries of hadron-production processes
$p+ \vec p \rightarrow h +X$.
These distributions are spin-dependent TMD distributions,
namely correlations between the transverse momentum and spin,
so that they should be related to angular momentum effects
in the nucleon.
Their measurements are valuable for understanding transverse-spin
structure and for testing an interesting relation to semi-inclusive
lepton measurement in the sense the TMD distributions change sign
due to gauge-link properties in these processes \cite{p24}. 
An advantage of J-PARC measurements is, for example, illustrated
in Fig. \ref{fig:ssa-d}, where single spin asymmetries
of $D$-meson production at J-PARC and RHIC are shown by considering
the Sivers' mechanism \cite{j-parc-d}. It is obvious that quark Sivers
functions are determined well at J-PARC, whereas measurements are
sensitive to gluon Sivers functions at RHIC. Therefore, the J-PARC
experiment is complementary to the RHIC one.

\vspace{-0.3cm}
\subsubsection{
Applicability of perturbative QCD to elastic single-spin asymmetry \\}

$ \ \ \ $ \vspace{-0.4cm}

\begin{wrapfigure}{r}{0.34\textwidth}
   \vspace{-0.5cm}
\begin{center}
   \includegraphics[width=0.29\textwidth]{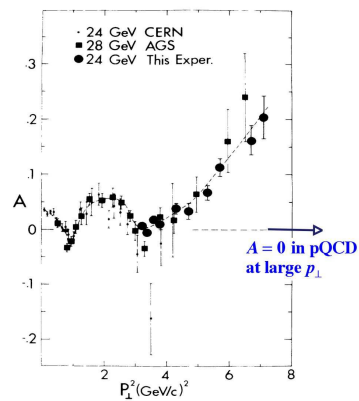}
\end{center}
    \vspace{-0.5cm}
\caption{Elastic single spin asymmetry in $p \vec p$ 
         \cite{elastic-spin}.}
\label{fig:elasic-spin}
\end{wrapfigure}

The perturbative QCD has been established for many high-energy
processes. However, its applicability to elastic spin asymmetries
is not understood in the sense, for example, that the single-spin
asymmetry in $p \vec p$ tends to increase as $p_\perp$ becomes larger
according to AGS measurements \cite{elastic-spin}, whereas it should
vanish in perturbative QCD. As shown in Fig. \ref{fig:elasic-spin},
the AGS measurements are up to $p_\perp^2 =8$ GeV$^2$, 
where some nonperturbative mechanism would still contribute
to the finite asymmetry. In any case, the AGS measurements need
to be confirmed by an independent experiment because it is 
difficult to find a possible mechanism to explain it. 
However, the current 30-GeV beam energy is similar to the AGS one,
so that other observables need to be considered in addition
to a mere confirmation, for example, by changing targets
and/or by considering angular distributions in order to provide
clues for theorists to understand the mechanism for the finite
asymmetry.

\vspace{-0.3cm}
\subsubsection{
Generalized parton distributions at hadron facilities \\} 

$ \ \ \ $ \vspace{-0.4cm}

\begin{wrapfigure}{r}{0.35\textwidth}
   \vspace{-1.4cm}
\begin{center}
   \includegraphics[width=0.28\textwidth]{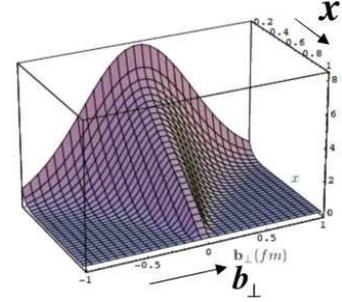}
\end{center}
    \vspace{-0.5cm}
\caption{3-dimensional picture of nucleon by GPDs \cite{gpd-3d}.}
\label{fig:gpd-3d}
\end{wrapfigure}

\noindent
Generalized parton distributions (GPDs) are global quantities 
in describing the nucleon because their forward limits are form
factors, first moments are the PDFs, and second moments are
related to orbital angular momentum contributions to the nucleon spin. 
In Fig. \ref{fig:gpd-3d}, the GPD integrated over the transverse-momentum
transfer
$ H(x,\vec b_\perp) = 
  \int \frac{d^2 \Delta_\perp}{(2\pi)^2} 
  e^{i \vec b_\perp \cdot \vec\Delta_\perp} 
  H(x,\xi=0,-\vec\Delta_\perp^2)$,
where $\xi$ is the skewdness parameter \cite{gpd-3d,gpd-hadron},
is shown for illustrating that the GPDs are related to
the 3-dimensional picture of the nucleon.  
  
\begin{figure}[h!]
\begin{minipage}{0.38\textwidth}
   \vspace{-0.0cm}
\begin{center}
   \includegraphics[width=0.85\textwidth]{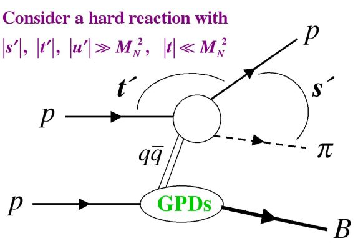}
\end{center}
   \vspace{-0.6cm}
\caption{GPD studies at hadron facilities \cite{gpd-hadron}.}
\label{fig:gpd-hadron}
\end{minipage}
\hspace{0.5cm}
\begin{minipage}{0.58\textwidth}
   \vspace{-0.4cm}
\begin{center}
   \includegraphics[width=0.95\textwidth]{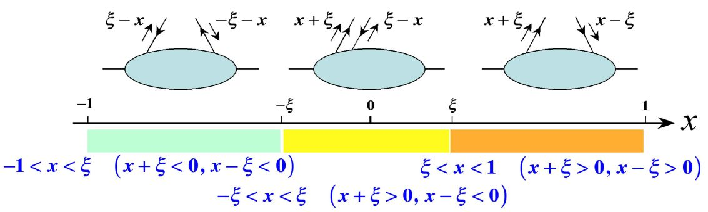}
\end{center}
\vspace{-0.3cm}
\caption{Three kinematical regions of GPDs \cite{gpd-hadron}.}
\label{fig:gpd-kinematics}
\end{minipage} 
\end{figure}

They have been studied mainly in virtual Compton process in
deep inelastic lepton scattering. However, the GPD studies could be
done at hadron facilities by using exclusive hadron-production
reactions $a+b \rightarrow c+d+e$ such as $N+N \rightarrow N + \pi + B$, 
where $B$ is $N$ (nucleon) or $\Delta$ as shown in Fig. \ref{fig:gpd-hadron}
\cite{gpd-hadron}.
Here, hadrons $c$ and $d$ have large and nearly opposite
transverse momenta and a large invariant energy, so that
an intermediate exchange could be considered as a $q\bar q$ state. 
The $q\bar q$ attached to the nucleon or $N \rightarrow \Delta$ is
expressed by the GPDs in a special kinematical region, so called 
Efremov-Radyushkin-Brodsky-Lepage (ERBL) region ($-\xi<x<\xi$),
which is in the middle of three regions of Fig. \ref{fig:gpd-kinematics}.
The hadron measurements of the GPDs are complementary to lepton-facility 
experiments in the sense that the specific kinematical region (ERBL)
is probed by extending the region of $x$ to medium $x$ because of
the J-PARC beam energy (30-50 GeV) and large cross sections of strong
interactions.

\vspace{-0.3cm}
\subsubsection{
Tensor structure in terms of quark and gluon degrees of freedom \\}

$ \ \ \ $ \vspace{-0.4cm}

\begin{wrapfigure}{r}{0.42\textwidth}
   \vspace{-0.3cm}
\begin{center}
   \includegraphics[width=0.30\textwidth]{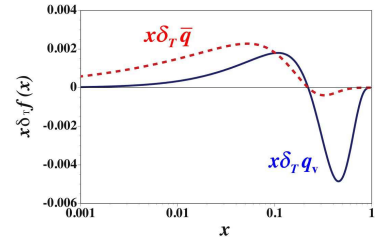}
\end{center}
    \vspace{-0.5cm}
\caption{Situation of tensor-polarized PDFs \cite{sk-b1}.}
\label{fig:tensor-b1}
\end{wrapfigure}

Tensor structure often appears in nuclear physics. For example,
the $D$-state admixture in the deuteron is the origin for
the finite electric quadrupole moment of the deuteron. 
It is interesting to describe the tensor structure
in terms of quark and gluon degrees of freedom.
Deep inelastic charged-lepton scattering from a tensor-polarized
deuteron was investigated by the HERMES collaboration, and
tensor polarized PDFs are extracted from their data \cite{sk-b1}.
It is particularly interesting to find a finite
tensor-polarized antiquark distribution $\delta_T \bar q (x)$,
which comes from the violation of the sum rule 
$\int dx b_1(x)=0$ \cite{sk-b1}
as shown in Fig. \ref{fig:tensor-b1}. 
This finite $\delta_T \bar q(x)$ can be directly
measured at hadron facilities by Drell-Yan processes 
with tensor-polarized deuteron ($p + \vec d \rightarrow \mu^+ \mu^- +X$)
\cite{pd-drell-yan}.
Here, the polarized-proton beam is not
needed for investigating this unique spin quantity.
The tensor structure has been investigated in the hadron degrees
of freedom; however, it begins to be understood in the parton level.

\subsection{Spin physics with proton-beam polarization}\label{spin-pol}

\begin{wrapfigure}{r}{0.45\textwidth}
   \vspace{-1.0cm}
\begin{center}
   \includegraphics[width=0.38\textwidth]{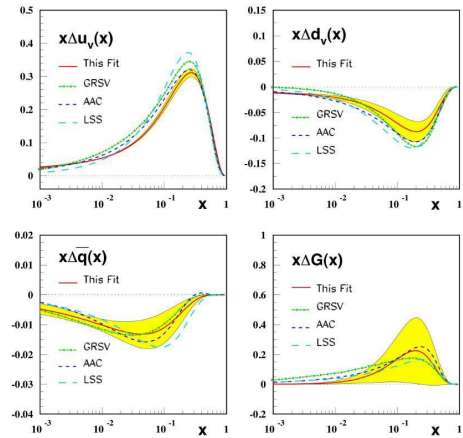}
\end{center}
    \vspace{-0.4cm}
\caption{Polarized PDFs \cite{pol-pdfs}.}
\label{fig:polpdfs}
\end{wrapfigure}

There is a proposal to polarize the primary proton beam \cite{p24}.
If it is attained, it becomes a complementary facility to the RHIC-Spin
project in the sense that the medium-$x$ region $0.2<x<0.7$ can be
investigated, whereas the RHIC probes a smaller-$x$ region.
For finding each partonic contribution to the nucleon spin,
the polarized PDFs should be integrated from 0 to 1,
at least from the kinematical region of RHIC to the one of J-PARC.
The J-PARC is a unique facility to investigate the polarized antiquark
distributions at relatively large $x$. At this stage, there is no reliable
flavor decomposition in the polarized antiquark distributions and
the polarized gluon distribution has not been determined
as shown in Fig. \ref{fig:polpdfs}.
For example, the flavor asymmetric distribution
$\Delta \bar u - \Delta \bar d$ has been measured by
the COMPASS collaboration \cite{compass-10};
however, the data are still not accurate enough to distinguish
various theoretical models. 

The transverse spin of the nucleon is another unsolved problem.
By measuring double spin asymmetries of Drell-Yan processes
with transversely polarized protons, we should be able to 
measure transversity distributions which are twist-two 
structure functions. Because of their chiral-odd property,
it does not couple to the gluon polarization,
which is unique and quite different from longitudinally-polarized
distributions. Therefore, the measurement of the transversity
provides us a valuable clue in understanding the nucleon spin.
At J-PARC with the proton-beam polarization, the transversity
distributions can be measured at medium $x$.

\subsection{High-energy hadron reactions in nuclear medium}\label{reaction}

Apart from the spin studies, there are interesting topics on
high-energy hadron reactions. Among them, we briefly discuss 
parton-energy loss, color transparency, and
short-range $NN$ correlations as examples of possible projects. 
These topics were partially studied at BNL-AGS and Fermilab, so that
J-PARC studies should be focused on new developments in this field.

\vspace{-0.3cm}
\subsubsection{Parton-energy loss \\}

$ \ \ \ $ \vspace{-0.4cm}

\begin{wrapfigure}{r}{0.34\textwidth}
   \vspace{-0.8cm}
\begin{center}
   \includegraphics[width=0.24\textwidth]{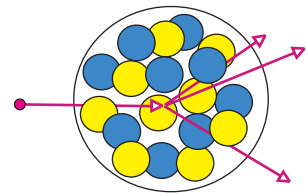}
\end{center}
    \vspace{-0.3cm}
\caption{Parton-energy loss.}
\label{fig:loss}
\end{wrapfigure}

Observation of jet suppression at large $p_T$ is one of
major results in heavy-ion collisions at RHIC. 
This phenomenon is considered to be an evidence of quark-gluon
plasma formation because a final-state parton loses energy when
it passes through the plasma. However, it is important to test
such a energy-loss mechanism in an ordinary nuclear medium instead
of the hot medium.

At J-PARC, the Drell-Yan process  $p+A \rightarrow \mu^+ \mu^- +X$
can be used for such a study \cite{p04} as illustrated in
Fig. \ref{fig:loss}. The muon pair is created by
the quark-antiquark annihilation ($q \bar q \rightarrow \mu^+  \mu^-$).
Therefore, the parton-energy loss in the initial quark should be
observed as an energy loss of the muon pair. Measurements could
be done for various energies and nuclear targets, so that their
dependencies should impose constraints on 
the energy-loss mechanism. This investigation affects
the studies on basic properties of quark-gluon plasma 
in heavy-ion reactions.

\vspace{-0.3cm}
\subsubsection{
Color transparency \\}

$ \ \ \ $ \vspace{-0.4cm}

\begin{wrapfigure}{r}{0.42\textwidth}
   \vspace{-1.2cm}
\begin{center}
   \includegraphics[width=0.40\textwidth]{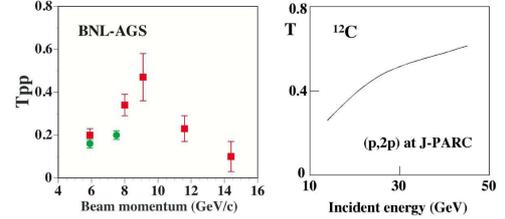}
\end{center}
    \vspace{-0.4cm}
\caption{Color transparency data of AGS and expected J-PARC
         results \cite{c-trans}.}
\label{fig:color-trans}
\end{wrapfigure}

Hadron interactions in nuclear medium are important for
understanding basic dynamical properties of QCD and for 
applications to other high-energy hadron interactions. 
At large momentum transfer, a small-size hadron is expected to
freely passes through nuclear medium. It is called color transparency,
which should be observed in high-energy hadron reactions.
Actual AGS measurements are shown in Fig. \ref{fig:color-trans},
where the ordinate is the nuclear transparency defined by
$T=\sigma_A/(A \sigma_N)$ with the cross section of 
$p A \rightarrow pp (A-1)$. As expected, the transparency
increases with increasing beam momentum. However, it suddenly
decreases at $p>10$ GeV/$c$, which is difficult to be interpreted
theoretically. At J-PARC, it is interesting to measure 
the transparency in this region and then to extend
the kinematical region to larger energies, where more 
transparency is expected theoretically as shown
Fig. \ref{fig:color-trans}.

\vspace{-0.3cm}
\subsubsection{
Short-range nucleon-nucleon correlations \\}

$ \ \ \ $ \vspace{-0.4cm}

\begin{wrapfigure}{r}{0.40\textwidth}
   \vspace{-1.2cm}
\begin{center}
   \includegraphics[width=0.38\textwidth]{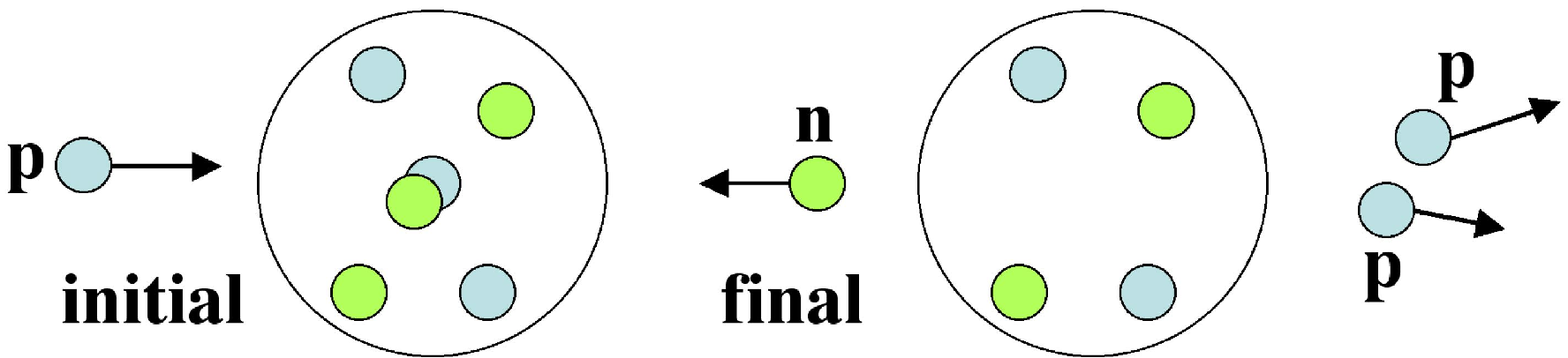}
\end{center}
    \vspace{-0.5cm}
\caption{$(p,2pN)$ process.}
\label{fig:short-range-p2pn}
    \vspace{+0.3cm}
\begin{center}
   \includegraphics[width=0.30\textwidth]{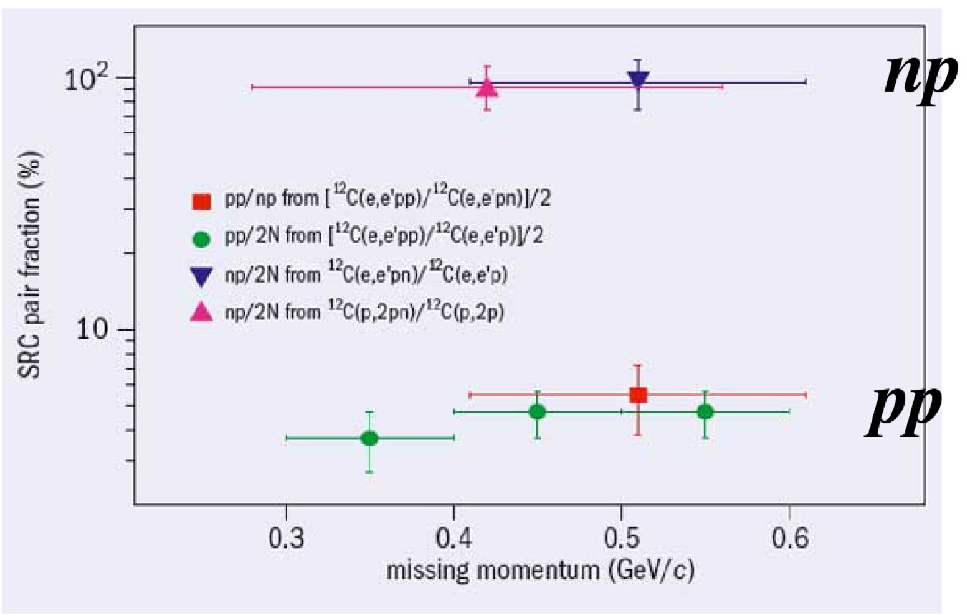}
\end{center}
    \vspace{-0.5cm}
\caption{Strong isospin dependence in short-range correlations 
         \cite{short}.}
\label{fig:short-range}
\end{wrapfigure}

Short-range correlations could be investigated at J-PARC,
for example, by the $(p,2pN)$ reaction as shown in 
Fig. \ref{fig:short-range-p2pn}.
Here, the 30-50 GeV proton incidents on a nuclear target
and two nucleons are emitted on opposite sides in the final state.
These nucleons are considered to be close together in
the initial state, so that the process is sensitive
short-range $NN$ interactions. Observing the reactions with
the $pp$ and $pn$ pairs in the final state, we can learn
about isospin dependence of the short-range correlations.
There are renewed interests in short-range $NN$ interactions,
where quark degrees of freedom should play an important role, 
due to recent developments in lattice QCD \cite{lattice-nn}.
This experimental project is attractive because of a recent finding
of strong isospin dependence, namely an unexpectedly large $np$
correlation over the $pp$ ($nn$) one.
Two-nucleon short-range correlations are experimentally
investigated in proton and electron reactions at BNL and JLab.
The results are surprizing in the sense that the $np$
correlation is much larger than the $pp$ one as shown in 
Fig. \ref{fig:short-range} \cite{short}. It could be explained
by the tensor force in the $NN$ interaction.
Such a difference between the $pn$ and $nn$ correlations could
affect neutron-star properties since a certain fraction of
protons exists in the stars. At J-PARC, it is valuable to study 
unexplored three-nucleon correlations at J-PARC 
in addition to the two-nucleon ones.

\section{Summary}\label{summary}

The J-PARC facility of the first stage has completed and 
hadron-physics experiments have started.
We introduced some of the major projects in this article
first on the approved experiments on strangeness hadron physics
and mass modifications in the nuclear medium, 
and second on future possibilities of hadron-structure physics.
In the beginning, new discoveries are expected for new hypernuclei,
hyperon interactions, kaonic nuclei, exotic hadrons with strangeness,
and mechanism of hadron-mass generation.
In the next stage, the various aspects of hadron structure could 
be investigated in addition. The possibilities of 
this second-stage projects were focused in this article.
The topics include flavor dependence 
of antiquark distributions, transverse-momentum-dependent distributions
such as the Boer-Mulders and Sivers functions, polarized exclusive reactions,
tensor structure functions, and generalized parton distributions.
Furthermore, high-energy parton and hadron reactions could be also 
investigated in nuclear medium. Among them, parton-energy loss,
color transparency, and short-range corrections were discussed.
These topics were introduced in this article just as examples
of future projects on hadron structure at J-PARC. 
Better ideas and actual proposals are essential
for the success of future J-PARC hadron project.

\vspace{-0.2cm}
\section*{Acknowledgements}
This work was partially supported by the Grant-in-Aid for Scientific
Research from the Japanese Ministry of Education, Culture, Sports,
Science, and Technology.

\vspace{-0.2cm}



\section*{References}
\begin{thebibliography}{9}
\bibitem{j-parc} 
               http://j-parc.jp/index-e.html;
               Akaishi Y, Hatsuda T, and Oka M 2002, 
               KEK Proceedings 2002-13 on JHF Nuclear Physics;
                 Kumano S 2007 {\it Nucl. Phys.} A {\bf 782} 442; 
                          2008 {\it AIP Conf. Proc.} {\bf 1056} 444.
\bibitem{inpc10-nagamiya-nagae}
         Nagamiya S, Nagae T 2010, talks at this conference,
         http://inpc2010.triumf.ca/.
\bibitem{sawada} Sawada S, 2007 {\it Nucl. Phys.} A {\bf 782} 434;
         talk at the conference,
         http://conferences.jlab.org/menu10/.
\bibitem{nuint09}
         For information on neutrino-nucleus interactions, 
         see http://nuint09.ifae.es/.
\bibitem{j-parc-proposal}
         For J-PARC proposals, 
         see http://j-parc.jp/NuclPart/Proposal\_e.html.
\bibitem{kaneta}
         Three-dimensional nuclear chart by M. Kaneta, Tohoku University.
\bibitem{neutron-star} 
         Schaffner-Bielich J 2008 {\it Nucl. Phys.} A {\bf 804} 309;
         Nishizaki S {\it et al.} 2002 {\it Prog. Theo. Phys.} {\bf 108} 703.
\bibitem{lattice-nn}
         Beane S R {\it et al.} 2006
                  {\it Phys. Rev. Lett.} {\bf 97} 012001; 
                  arXiv:1004.2935 [hep-lat] 
         Ishii N, Aoki S, and Hatsuda T 2007
                       {\it Phys. Rev. Lett.} {\bf 99} 022001;
         Yamazaki Y and Kuramashi Y 2010
                       {\it Phys. Rev.} D {\bf 81} 111504. 
\bibitem{exotic-project} See
      http://www.hepl.phys.nagoya-u.ac.jp/public/new\_hadron/index-e.html.
\bibitem{theta} 
        Nakano T {\it et al.} 2009
            {\it Phys. Rev.} C {\bf 79} 025210;
        J-PARC proposal P19, Naruki M {\it et al.} 2006.
\bibitem{s-theta} Letter of Intent P09-LoI, 
         Hotta T {\it et al.} 2006.
\bibitem{knnn} Akaishi Y {\it et al.} 2002
                  {\it Phys. Rev.} C {\bf 65} 044005;
                  2004 {\bf 70} 044313;
                  2005 {\it Phys. Lett.} B {\bf 613} 140.
\bibitem{kpp} J-PARC proposal P15, Iio M {\it et al.} 2006;
              P27, Nagae T {\it et al.} 2009.
\bibitem{k-he-x} 
         J-PARC proposal P17, Beer G {\it et al.} 2006.
\bibitem{lambda-1405} J-PARC proposal P31,
          Ajimura S {\it et al.} 2009.
\bibitem{hh-summary}
        Hayano R S and Hatsuda T, arXiv:0812.1702 [nucl-ex].
\bibitem{kek-ps} 
                 Naruki M {\it et al.} 2006
                     {\it Phys. Rev. Lett.} {\bf 96} 092301
                 [Figure 7 is printed with permission 
                 (\copyright 2006 American Physical Society (APS))];
                 Muto R {\it et al.} 2007
                     {\it loc. cit.} {\bf 98} 042501.
\bibitem{j-parc-workshop}
         KEK hadron workshops at
         http://www-conf.kek.jp/J-PARC-HS05/;
         http://www-conf.kek.jp/hadron08 \\ /hehp-jparc/;
         http://www-conf.kek.jp/hadron1/hehp-th10/.
\bibitem{resum} Shimizu H {\it et al.} 2005
                     {\it Phys. Rev.} D {\bf 71} 114007;
                Yokoya H and Vogelsang W 2007 
                     {\it AIP Conf. Proc.} {\bf 915} 595.
\bibitem{p04} J-PARC proposal P04, Chiba J {\it et al.} 2006.
\bibitem{p24} J-PARC proposal P24, Bai M {\it et al.} 2008.
\bibitem{flavor}  Kumano S 1998 {\it Phys. Rept.} {\bf 303} 183;
                  Garvey G T and Peng J-C 2001
                       {\it Prog. Part. Nucl. Phys.} {\bf 47}, 203.
\bibitem{hermes}
     The figure was provided by the HERMES collaboration, in
     particular U. Elschenbroich.
\bibitem{pol-pdfs} 
     Bl\"umlein J and B\"ottcher H 2010
             {\it Nucl. Phys.} B {\bf 841} 205
     [Figure 16 is obtained from J. Bl\"umlein].      
\bibitem{j-parc-d}
     Anselmino M {\it et al.} 2004
              {\it Phys. Rev.} D {\bf 70} 074025
     [Figure 10 is printed with permission (\copyright 2004 APS)];
      D'Alesio U and Murgia F 2007 
              {\it AIP Conf. Proc.} {\bf 915} 559. 
\bibitem{elastic-spin}
     Crabb D G {\it et al.} 1990 
              {\it Phys. Rev. Lett.} {\bf 65} 3241
     [Figure 11 is printed with permission (\copyright 1990 APS)].
\bibitem{gpd-3d}
      Burkardt M 2003 {\it Int. J. Mod. Phys.} A {\bf 18} 173
      [Figure 12 is reproduced with permission
       (\copyright 2003 World Scientific Publishing Company)].
\bibitem{gpd-hadron} Kumano S, Strikman M, and Sudoh K 2009
                     {\it Phys. Rev.} D {\bf 80} 074003. 
\bibitem{sk-b1} 
      Airapetian A {\it et al.} (HERMES Collaboration) 
               2005 {\it Phys. Rev. Lett.} {\bf 95} 242001;
      Kumano S 2010 {\it Phys. Rev.} D {\bf 82} 017501;
      Close F E and Kumano S 1990 {\it Phys. Rev.} D  {\bf 42} 2377.
\bibitem{pd-drell-yan}
      Hino S and Kumano S 1999 {\it Phys. Rev.} D {\bf 59} 094026;
                                                  {\bf 60} 054018.
\bibitem{compass-10} 
      Alekseev M G {\it et al.} (COMPASS Collaboration) 2010
      arXiv:1007.4061 [hep-ex].
\bibitem{c-trans} 
       Aclander J {\it et al.} 2004  
              {\it Phys. Rev.} C {\bf 70}  015208;
       Strikman M 2005, talk in Ref. \cite{j-parc-workshop};
       Kohama A, Yazaki K, and Seki R 1992,
              {\it Nucl. Phys.} A {\bf 536} 716. 
\bibitem{short} Higinbotham D, Piasetzky E, and Strikman M 2009
                    {\it CERN Courier} {\bf 49} 22;
                Piasetzky E {\it et al.} 2006 
                    {\it Phys. Rev. Lett.} {\bf 97} 162504;
                Alvioli M, Ciofi degli Atti C, and Morita H 2008 
                    {\it Phys. Rev. Lett.} {\bf 100} 162503. 
\end{thebibliography}
\end{document}